\documentclass[12pt,draft]{article}
  \usepackage{amsthm}
  \usepackage{amssymb}
  \usepackage{amsmath}
  \usepackage{eucal}
  \usepackage{mathrsfs}
  \usepackage[all]{xy}
 \usepackage{graphicx}
  \usepackage{bbm}

\newcommand{\D}{\mathrm{D}}
\newcommand {\norm}[1]{\Vert{#1}\Vert}

\newcommand{\R}{\mathbb{R}} 
\newcommand{\Hil}{\mathcal{H}} 

\newcommand{\pa}{\partial}

\newcommand{\scp}[2]{\left\langle #1,#2\right\rangle}
\newcommand{\bra}[1]{\left\langle #1\right\vert}
\newcommand{\ket}[1]{\left\vert #1 \right\rangle}

 \author{Romeo Brunetti$^{(1)}$, Klaus Fredenhagen$^{(2)}$ and Marc Hoge$^{(2)}$ \\
  \null\\
  \small{$^{(1)}$ Dipartimento di Matematica, Universit\`a di Trento,}\\
  \small{Via Sommarive 14, I-38123 Povo (TN), Italy}\\[10pt]
  \small{$^{(2)}$ II. Institut f\"ur Theoretische Physik, Universit\"at Hamburg,}\\
    \small{Luruper Chaussee 149, D-22761 Hamburg, Germany}\\ [10pt]
  \small{\texttt{romeo.brunetti@gmail.com, klaus.fredenhagen@desy.de}}\\
  \small{\texttt{marc.hoge@desy.de}}}

\title{Time in quantum physics: From an external parameter to an intrinsic observable}
\date{}
 
\begin{document}

  \maketitle

\begin{abstract}
In the Schr\"odinger equation, time plays a special role as an external parameter. We show that in an enlarged system where the time variable denotes an additional degree of freedom, solutions of the 
Schr\"odinger equation give rise to weights on the enlarged algebra of observables. States in the associated GNS representation correspond to states on the original algebra composed with a completely positive unit preserving map. Application of this map to the functions of the time operator on the large system delivers the positive operator valued maps which were previously proposed by two of us as time observables. As an example we discuss the application of this formalism to the Wheeler-DeWitt theory of a scalar field on a Robertson-Walker spacetime.
\end{abstract}
  \theoremstyle{plain}
  \newtheorem{df}{Definition}[section]
  \newtheorem{teo}[df]{Theorem}
  \newtheorem{prop}[df]{Proposition}
  \newtheorem{cor}[df]{Corollary}
  \newtheorem{lemma}[df]{Lemma}

  \theoremstyle{definition}
  \newtheorem{oss}[df]{Remark}

\theoremstyle{definition}
  \newtheorem{ass}{\underline{\textit{Assumption}}}[section]
\section{Introduction}
The main conceptual problem in quantum gravity is that spacetime which serves as a mean for parametrizing quantum fields must  itself be treated in terms of  observables in the sense of quantum physics. A plausible first attempt is to replace spacetime by a noncommutative space. One then has to guess the structure of this space and to understand the way physics can be described on a noncommutative background.

In spite of the big effort which was put into this program, the results which have been obtained so far are, in our opinion, not really satisfactory. On the conceptual level, there is  a large arbitrariness in the choice of the noncommutative structure; only uncertainty relations have been predicted with some plausibility \cite{DFR}. Even more, the whole idea of putting physics on a fixed background seems to exclude dynamical effects of quantum gravity. On the practical level it turned out to be extremely difficult to put interacting theories on a noncommutative background. The best results which are available seem to be the results of Grosse and Wulkenhaar \cite{GW}. But the model they construct is very far from relativistic quantum physics and even more from quantum gravity.

One of the most popular approaches to quantum gravity is the Ashtekar program, in which one tries to quantize gravity directly. Most successful in this approach is the version of loop quantum gravity where, however, the emergence of a classical spacetime in an appropriate limit remains a challenge. Recent progress was obtained by the introduction of so-called relational observables \cite{Rovelli,Dittrich,Thiemann}. The basic idea is that the unphysical parameter spacetime is eliminated from the description by expressing fields as functions of other fields. In classical physics this amounts to an application of the theorem on implicit functions, and, up to possible zeros of the corresponding Jacobian, the program seems to work. In the quantum case, however, the program leads to difficult questions.

In a simplified situation the problem has been studied by Ashtekar, Paw\-low\-ski, and Singh for Robertson-Walker spacetimes, where the scale parameter and a spatially constant scalar field are the degrees of freedom \cite{Ashtekar}. The associated field equation can be understood as a Schr\"odinger equation with energy zero. The program then is to reinterpret the solution as a time dependent Schr\"odinger equation where the scalar field plays the role of the time parameter. 

In this article we want to discuss a similar problem in standard quantum mechanics. We will apply this program to the problem of defining time as an observable, and we will compare the result with the construction of time observables two of us did before \cite{BFtime}. Finally, a novel analysis of the scalar field on a Robertson-Walker spacetime is given which uses the techniques developed in this paper.

\section{Schr\"odinger solutions as weight on the algebra of observables}

Let us look at the Schr\"odinger equation
\begin{equation}
i\frac{d}{dt}\psi(t)=H_0\psi(t) \ ,
\end{equation}      
where $H_0$ is a selfadjoint operator on a Hilbert space $\Hil_0$ and $\psi$ is a continuously differentiable function on $\R$ with values in the domain of $H_0$. 

It is tempting to write the Schr\"odinger equation in the form
 \begin{equation}
 H\psi=0\ ,
\end{equation}  
where 
\begin{equation}
H:=\frac{1}{i}\frac{d}{dt}+H_0 \ ,
\end{equation}  
is a selfadjoint operator on 
\begin{equation}
\Hil :=L^2(\R,\Hil_0) \ .
\end{equation}  
The difficulty with this (well known) idea is that the spectrum of $H$ is continuous, hence $\psi\not\in\Hil$.
Traditionally, one interprets solutions of the eigenvalue equation for points in the continuous spectrum as linear functionals on a suitable dense domain $\D\subset\Hil$. 
But there is another possibility which has to our knowledge in physics first been applied by Buchholz, Porrmann, and Stein in their approach to the infrared problem \cite{BP,P}. In this work they interpret improper states with sharp momentum as weights, i.e. positive linear functionals on the algebra of observables which are finite only on a suitable subset. The trace on an infinite dimensional Hilbert space is an example. (cf. \cite{BR} for some mathematical background.)  

Actually, if $\psi$ is a linear functional on $\D$, we can define the associated weight $w_{\psi}$ on positive operators in the following way. 
Let $A$ be a bounded operator with $A\Hil\subset \D$ such that $\psi\circ A$ is a continuous linear functional on $\Hil$. Examples of such operators are finite sums of rank 1 operators of the form $\ket{\chi}\bra{\varphi}$ with $\chi\in \D$ and $\varphi\in\Hil$. Then the adjoint of $A$ acts on $\psi$ by
\begin{equation}
(A^*\psi)(\varphi)=\psi(A\varphi) \ , \ \varphi\in\Hil \ ,
\end{equation}
and $A^*\psi$ can be identified with an element of $\Hil$. The weight can then be defined on operators of the form $AA^*$ by
\begin{equation}
w_{\psi}(AA^*)=||A^*\psi||^2\ .
\end{equation}
We then extend the definition to the additive cone $C_{\D}$ generated by these operators by requiring additivity of the weight. In a last step we set for an arbitrary positive operator
\begin{equation}
w_{\psi}(B)=\sup_{0\le A\le B,\ A\in C_{\D}}w_{\psi}(A) \ .
\end{equation}

The weight has no direct interpretation in terms of probability distributions since it is not normalizable.
It has, however, an interpretation in terms of conditional probabilities. Namely, 
let us consider the GNS representation $(\pi_{\psi},\Hil_{\psi})$ associated to the weight. It is constructed in terms of the left ideal
\begin{equation}
L_{\psi}=\{A\in B(\Hil)|w_{\psi}(A^*A)<\infty\} \ .
\end{equation}
By the polarization identity we extend the weight to the algebra $L_{\psi}^*L_{\psi}$. We then equip
$L_{\psi}$ with the positive semidefinite scalar product
\begin{equation}
(A,B)=w_{\psi}(A^*B) \ .
\end{equation}
$\Hil_{\psi}$ is the completion of the pre-Hilbert space obtained by dividing out the null space of the scalar product. The representation $\pi_{\psi}$ is induced by left multiplication. We may interpret the state associated to $A\in L_{\psi}$ as the association of probabilities under the condition that the effect $A^*A$ took place. Note that the conditional probabilities defined in this way depend not only on the effect itself, but also on the way it is written as an absolute square, i.e. on the phase of $A$. Quantum mechanically this may be understood as the dependence of the state on the way the event $A^*A$ influenced the state. 

We apply this general construction now to the solutions $\psi$ of the time dependent Schr\"odinger equation. The left ideal contains in particular the operators of multiplication by bounded square integrable functions $g$ of $t$. The corresponding states may be restricted to $B(\Hil_0)$.  
We find
\begin{equation}
\omega_{g\psi}(\cdot)=\frac{\int dt\,|g(t)|^2\scp{\psi(t)}{\cdot \ \psi(t)}}{\int dt \,|g(t)|^2\norm{\psi(t)}^2} \label{eq:condprob}\ .
\end{equation}
In the limit when $|g(t)|^2 dt$ approaches the Dirac measure at the point $t_0$, we recover the state induced by $\psi(t_0)\in\Hil_0$. Moreover, the time evolution of the large system acts as a translation for $g$ and thus induces the time evolution in the small system. 
We see that the new formalism covers the standard formalism of quantum mechanics. It also nicely reproduces the standard interpretation of states in the Schr\"odinger picture saying that $\psi(t_0)$ is the state under the condition that the time observable of the enlarged system takes the value $t_0$.

There are in general also other elements in the left ideal $L_{\psi}$. Namely, an element $A\in B(\Hil_0)$ being in the left ideal means that the integral
\[
\int dt \scp{\psi(0)}{e^{iH_0t}A^*Ae^{-iH_0t}\psi(0)}<\infty
\] 
is finite. If the spectrum of $H_0$ is absolutely continuous, there are many operators with this property.
 The value of the integral may be interpreted as the time duration of the event $A^*A$.  
 
Let $\tau_A$ be the unbounded positive selfadjoint operator whose expectation values are given by these integrals. We restrict ourselves to the case when  $\tau_A$ has trivial kernel. Since $\tau_A$ commutes with the Hamiltonian $H_0$, we can always reach this situation by replacing $\Hil_0$ by the orthogonal complement of the kernel of $\tau_A$.

The states $\omega_{A\psi}$ on $B(\Hil)$ and $\omega_{\sqrt{\tau_A}\psi(0)}$ on $B(\Hil_0)$ are related by
\begin{equation}
\omega_{A\psi}=\omega_{\sqrt{\tau_A}\psi(0)}\circ\Phi_A \ ,
\end{equation}   
with the completely positive unit preserving map $\Phi_A:B(\Hil)\to B(\Hil_0)$ which is given by
\begin{equation}
\Phi_A(C)=V_A^*CV_A \label{eq:posmap}
\end{equation}
where the isometry 
\[V_A:\Hil_0\to \Hil=L^2(\R,\Hil_0)\]
is defined by
\[
(V_A\chi)(t)=Ae^{-itH_0}\tau_A^{-\frac12} \chi \ ,\ \chi\in\Hil_0 \ .
\] 

Using this positive mapping, one may relate observables of the enlarged system to observables of the small system. In particular, one may apply it to the time observable. The time parameter $t$ of the Schr\"odinger equation is a selfadjoint multiplication operator $T$ on the enlarged Hilbert space $\Hil = L^2(\R, \Hil_0)$. Its spectral projections $E_T(I)$, $I\subset\R$ measurable, can be mapped to positive operators $\Phi_A(E_T(I))$, and one obtains the positive operator valued measure (POVM)
\begin{equation}
I \quad \mapsto \quad \Phi_A (E_T(I)) = \tau_A^{-\frac12} \ \int_I dt \, e^{iH_0 t} \, A^\ast A \, e^{-iH_0 t} \, \tau_A^{-\frac12} \ . \label{eq:povmbf}
\end{equation}
Since the spectral projections $E_T(I)$, $I\subset\R$ commute with $A$, the POVM does not depend on the phase of $A$.

The POVM above was first constructed  in \cite{BFtime} and there, in agreement with the more complete discussion of the present paper, interpreted as the time of occurrence of the event $A^\ast A$.

The present framework is closely related to the concept of partial and complete observables initiated by Rovelli \cite{Rovelli} and the theory of constraints. The observables of our enlarged Hilbert space are the partial observables in the sense of Rovelli. Complete observables are those partial observables which are invariant under the automorphism induced by the Hamiltonian $H$ of the enlarged system. They are the Dirac observables if $H$ is considered as a constraint. As emphasized by Rovelli, the partial observables themselves have an observable meaning since relations between them give rise to complete observables. In the present framework these relations take the form of conditional probabilities.

Contrary to the formalism of Rovelli, the partial observables in our framework are well defined operators which act on the GNS Hilbert space associated to the weight. It is also clear from our construction that one should not expect, without additional structure, a unique state on the Dirac observables in case of a constraint operator with continuous spectrum, since the left ideal $L_{\psi}$ will not contain the unit operator.  

Conditional probabilities in quantum physics and their use for an introduction of time observables have also been discussed by Gambini, Pullin and coll. \cite{GP}. The importance of the concept of weights for a proper definition in the case of constraints with continuous spectrum was apparently overlooked because of the emphasis on the discrete case. 
\section{Nonrelativistic motion in 1 dimension}
The simplest example on which we can apply our formalism is a nonrelativistic free particle on a line.
Let $\Hil_0=\mathrm{L}^2(\R)$, $\Hil=\mathrm{L}^2(\R^2)$. The Hamiltonian $H_0$ on $\Hil_0$ is
\[
H_0=-\frac{1}{2m}\frac{d^2}{dx^2}
\]  
the Hamiltonian in the enlarged formalism is
\[
H=\frac{1}{i}\frac{\partial}{\partial t}-\frac{1}{2m}\frac{\partial^2}{\partial x^2} \ .
\]
We ask the question: when is the particle at the point $x_0$? This question may be rephrased in the form: what is the value of time under the condition that the particle is at the point $x_0$? 
According to the discussion in the previous section, the probability that the time observable is within the interval $I$ under the condition that the position of the particle is within the space interval $J$ is
\begin{equation} \label{P(I|J)}
P(I|J)=\frac{w_{\psi}(E_X(J)E_T(I)E_X(J))}{w_{\psi}(E_X(J))}
\end{equation}
where $E_T(I)$ is the spectral projection of the time observable on $L^2(\R^2)$, and
$E_X(J)$ the spectral projection of the space observable on $L^2(\R^2)$.

A solution of the Schr\"odinger equation is of the form
\[
\psi(t,x)=\int dp e^{-i\frac{p^2}{2m}t+ipx}\varphi(p) \ .
\]
with a square integrable function $\varphi$.
The numerator in equation (\ref{P(I|J)}) is given by
\begin{align}
w_{\psi}(E_X(J)E_T(I)E_X(J))&=w_{\psi}(E_X(J)E_T(I))\notag\\
&=\int_I dt\int_Jdx\int dp\int dq\, e^{\frac{it}{2m}(p^2-q^2)-i(p-q)x}\overline{\varphi(p)}\varphi(q) \ ,
\end{align}
and for the normalization factor in equation (\ref{P(I|J)}) we obtain
\begin{align}w_{\psi}(E_X(J))&=\int_{-\infty}^{\infty} dt\int_Jdx |\psi(t,x)|^2\notag\\
&=\int_Jdx\int dp\int dq\, 2\pi \delta(\frac{p^2-q^2}{2m})e^{-i(p-q)x}\overline{\varphi(p)}\varphi(q)\notag\\
&=2\pi \int_Jdx\int dp
\frac{m}{|p|}\overline{\varphi(p)}\left(\varphi(p)+e^{-2ipx}\varphi(-p)\right)
\end{align}
This factor is finite if $\varphi$ is smooth and vanishes at $p=0$. Hence for such solutions $\psi$ of the Schr\"odinger equation the spectral projections $E_X(J)$ are elements of the left ideal $L_{\psi}$. Note that for large intervals $J$ it approaches the value $|J|\langle\frac{1}{v}\rangle$ with the expectation value of the inverse velocity $\frac{1}{v}$, in agreement with the classical amount of time the particle spends inside the interval ($|J|$ is the length of the interval). 

Let $J=[-a/2,a/2]$, and let $\varphi=\varphi_{+}+\varphi_{-}$ be the decomposition of $\varphi$ into the even and the odd part. 
Then the completely positive mapping associated to $E_X(J)$ is given by
\begin{equation}
\phi_a(A)=V_a^*AV_a
\end{equation}
where the isometry $V_a:L^2(\R)\to L^2(\R^2)$ acts on $\psi_0(x)=\int dp e^{ipx}\varphi(p)$ by
\[
(V_a\psi_0)(t,x)=\int dp\, e^{-it\frac{p^2}{2m}+ipx}\sqrt{\frac{|p|}{m}}\sum_{\pm}\left(a\pm\frac{\sin a p}{ p}\right)^{-\frac12}\varphi_{\pm}(p)
\]
if $x\in J$ and $V_a\psi_0(t,x)=0$, otherwise.

In the limit $a\to 0$ the POVM's $P_a(I)=\phi_a(E_T(I))$ converge, and the POVM $P_0$ which describes the probability that the particle reaches the origin has the momentum space integral kernel
\begin{equation}
P_0(I)(p,q)=\left\{
\begin{array}{ccc}
\frac{\sqrt{pq}}{2\pi m}\int_I dt\, e^{it\frac{p^2-q^2}{2m}} & , & pq>0 \\
                               0                                              & , & \text{else}\ .
\end{array}  \right.\notag                                       
\end{equation}
The first moment of this measure (the associated time operator on $\Hil_0$) turns out to coincide with
Aharanov-Bohm's time operator \cite{Aharanov} 
\begin{equation}
T=-\frac{m}{2}(p^{-1}x+xp^{-1})
\end{equation}
Note that $T$ is not selfadjoint, but maximally symmetric with deficiency indices $(2,0)$. (cf. \cite{Busch}).
\section{Relativistic motion in 1 dimension}
We may now redo the analysis for a relativistic particle. Its motion is described by a positive frequency solution $\psi$ of the Klein Gordon equation. In analogy to the previous case we look at the operator 
$\frac{\pa^2}{\pa t^2}-\frac{\pa^2}{\pa x^2}+m^2$ on $\mathrm{L}^2(\R^2)$. $\psi$ has the form
\[
\psi(t,x)=\int dp\, e^{-i\sqrt{p^2+m^2}t+ipx}\varphi(p) \ .
\]
The calculations are completely analogous to the nonrelativistic case. One replaces the energy $\frac{p^2}{2m}$ by the relativistic energy $\sqrt{p^2+m^2}$ and the velocity $\frac{p}{m}$ by the relativistic velocity $\frac{p}{\sqrt{p^2+m^2}}$.
Thus one finds for the POVM $P_0$ 
\begin{equation}
P_0(I)(p,q)=\left\{
\begin{array}{ccc}
\frac{1}{2\pi}\sqrt{\frac{pq}{\sqrt{p^2+m^2}\sqrt{q^2+m^2}}}\int_I dt\, e^{it(\sqrt{p^2+m^2}-\sqrt{q^2+m^2})} & , & pq>0 \\
                               0                                              & , & \text{else}.
\end{array}  \right.                                       
\end{equation}
The relativistic analogue of Aharonov-Bohm's time operator is
\begin{equation}
T=-\frac12\left(\frac{\sqrt{p^2+m^2}}{p}x+x\frac{\sqrt{p^2+m^2}}{p}\right) \ .
\end{equation}

\section{Application to a cosmological model}

The formalism on an enlarged Hilbert space can be applied to the afore mentioned analysis of a massless scalar field on a Robertson-Walker spacetime carried out by Ashtekar, Pawlowski and Singh \cite{Ashtekar} (see also \cite{AshtekarKL}).

The classical phase space of the model is coordinatized by $(c,p \, ; \phi, p_\phi)$ where $\phi$ is the scalar field and $p_\phi$ its conjugate momentum. The pair $(c,p)$ represents the gravitational degree of freedom with $p$ being proportional to the square of the scale factor. Hence, $p$ is restricted to non-negative values. The non-trivial Poisson brackets are given by
\[
\{ c,p\} = \frac{8\pi \gamma G}{3} \qquad \text{and} \qquad \{\phi,p_\phi\} = 1  
\]
where $\gamma$ is the Barbero-Immirzi parameter and $G$ the Newton constant. Along the classical trajectories $\phi$ is a monotonous function and can therefore be used as a ``time'' parameter. The goal of this section is to use $\phi$ as an ``emergent time'' also in quantum theory. Following \cite{Ashtekar} the Wheeler-DeWitt equation of the model can be written as
\begin{equation}
\left(\frac{\pa^2}{\pa \phi^2} + \Theta \right) \Psi(\phi,p) = 0 \ , \quad \Theta := - \frac{16\pi G}{3} \, p^{\frac{3}{2}}\, \frac{\pa}{\pa p}\sqrt{p}\frac{\pa}{\pa p} \ . \label{eq:wdw}
\end{equation}
By the unitary transformation $V:L^2(\R,du)\to\Hil_0$, 
\begin{equation}
(V\phi)(p)=p^{\frac14}\phi(\ln p)
\end{equation}
the operator $\Theta$ gets the standard form
\begin{equation}
V^*\Theta V=\frac{16\pi G}{3}\left(-\frac{\pa^2}{\pa u^2}+\frac{1}{16}\right) \ .
\end{equation}
The dual variable $c$ can be represented by the operator
\[c= \frac{8\pi \gamma}{3i}e^{-\frac{u}{2}}\frac{\pa}{\pa u}e^{-\frac{u}{2}} \ .\]
The Wheeler-De Witt equation of the model is equivalent to the Klein Gordon equation.
We consider solutions of the form
\begin{equation}
\psi(\phi,u)=(2\pi)^{-1}\int d\omega  dk \, \delta\left(\omega^2-\kappa^2(k^2+\frac{1}{16})\right)e^{-i(\omega \phi-ku)}\varphi(\omega,k)
\end{equation}
with a test function of compact support $\varphi$ and the abbreviation 
\[\kappa := \sqrt{16 \pi G /3}.\] 
$\psi$ induces a weight on $B(L^2(\R^2))$ which is finite on bounded operators $A$ with a distributional integral kernel $a$ with compact support,
\begin{equation}
w_{\psi}(A)=a(\overline{\psi}\otimes\psi)\equiv\int d\phi du d\phi' du' \overline{\psi(\phi,u)}a(\phi,u;\phi',u')\psi(\phi',u') \ .
\end{equation}  
Let $\Phi$ be multiplication by $\phi$ and $U$ multiplication by $u$. The product of the corresponding spectral projections $E_{\Phi}(I)E_{U}(J)$ for finite intervals $I,J$ of the real line are in the domain of $w_{\psi}$. 

Also the projections $E_{\Phi}(I)$ are in the domain. We may actually perform the limit $I\to\{0\}$ and obtain
\begin{align*}
&\phantom{Platz Platz}\lim \frac{1}{|I|}w_{\psi}(E_{\Phi}(I))=\int du |\psi(0,u)|^2\\
&=(2\pi)^{-2}\int du d\omega  d\omega'  dk dk' \delta\left(\omega^2-\kappa^2(k^2+\frac{1}{16})\right)\delta\left(\omega'{}^2-\kappa^2(k'{}^2+\frac{1}{16})\right)\\ 
&\phantom{(2\pi)^{-2}\int du d\omega  d\omega'  dk dk' \delta\left(\omega^2-\kappa^2(k^2+\frac{1}{16})\right)}  
\times e^{-i(k-k')u}\overline{\varphi(\omega,k)}\varphi(\omega',k')\\
&=(2\pi)^{-1}\int d\omega d\omega' dk \delta\left(\omega^2-\kappa^2(k^2+\frac{1}{16})\right)\delta\left(\omega'{}^2-\omega^2\right)\overline{\varphi(\omega,k)}\varphi(\omega',k')\\
&=\sum_{\pm}\int dk |\varphi_{\pm}(k)|^2 =:\scp{\varphi}{\varphi}
\end{align*}
with 
\begin{equation}
\varphi_{\pm}(k)=(2\pi)^{-\frac12}\frac{\varphi(\pm\omega_k,k)}{2\omega_k}\ , \ \omega_k=\kappa\sqrt{k^2+\frac{1}{16}} \ .
\end{equation}
This corresponds to the state ``at time zero'' ($\phi=0$) as in the quantum mechanical discussion in Section 2.

We want to to check whether also $E_U(J)$ is in the domain of the weight. According to the interpretation given before, this would mean that the expected time duration (measured by $\phi$) for the scale parameter $p=e^u$ being in the interval $e^{J}$ is finite.
 
We compute
\begin{align*}
&w_{\psi}(E_{U}(J))=\int_J du \int_{\R}d\phi |\psi(\phi,u)|^2\\
&=2\pi \int_J du\int d\omega dk dk'\, \delta\left(\omega^2-\kappa^2(k^2+\frac{1}{16})\right)\delta(k^2-k'{}^2)e^{-i(k-k')u}\overline{\varphi(\omega,k)}\varphi(\omega,k')\\
&=2\pi \int_J du\int d\omega dk\, \delta\left(\omega^2-\kappa^2(k^2+\frac{1}{16})\right)\frac{1}{2|k|}\overline{\varphi(\omega,k)}\left(\varphi(\omega,k)+e^{-2iku}\varphi(\omega,-k)\right)\\
&= \scp{\varphi}{\frac{\omega_k}{|k|}(|J|+\sqrt{2\pi}\hat{\chi}_J(2k)\Pi)\varphi}
\end{align*}    
where $\chi_J$ is the characteristic function of the interval $J$ and $\Pi$ is the parity operator 
\[\Pi\varphi(\omega,k)=\varphi(\omega,-k) \ .\]
We find that the expected time duration is finite if $\varphi$ vanishes at $k=0$.

Under this condition we can now define the probability distribution of instants of time at which the scale parameter $p$ is inside the prescribed interval. The discussion is essentially the same as the discussion for the relativistic particle, the only difference being that we may decompose the solution $\psi$ in parts corresponding to expanding universes ($\frac{k}{\omega}>0$) and contracting universes ($\frac{k}{\omega}<0)$ instead of a decomposition with respect to the sign of the frequency. For the time operator on the time zero states which characterizes the instant of time, when the  parameter $p=e^u$ is equal to 1, we find
\begin{equation}
T_1=\frac12\left(\frac{\omega}{k}u+u\frac{\omega}{k}\right) \ .
\end{equation}   
Here $(\omega\varphi)_{\pm}=\pm\omega_k\varphi_{\pm}$ and $u\varphi(\omega,k)=i\frac{\pa}{\pa k}\varphi(\omega,k)$.

The time operator for other values of $p$ follows by covariance,
we obtain
\begin{equation}
T_p=e^{ik\ln p}T_0 e^{-ik\ln p}=T_0+\frac{\omega}{k}\ln p \ .
\end{equation}
We conclude that expanding universes reach the big bang ($p=0$) at $\phi=-{\infty}$. 
\section{Summary}
In the present paper we proposed an extended interpretation of quantum mechanics, based on the 
operator algebraic concept of weights and their interpretation in terms of conditional probabilities. 
In this formalism, the Schr\"odinger picture arises as an association of conditional probabilities where the condition is that the time variable assumes a given value. The formalism allows also other questions, in particular: 
What is the value of time when a given event happens? 
It is shown that this concept of an observable time reproduces the time variables proposed by two of 
us some time ago, and we illustrated the concept on the example of the Aharanov-Bohm time operator, 
its relativistic generalization and a cosmological time observable in the Wheeler-De Witt model of a 
quantized Robertson-Walker spacetime. 

The general concepts of this paper might be used also for a new treatment of constraints, in particular in the case when the constraints have continuous spectrum. Our ideas are related to Rovelli's concept of partial observables and the Gambini-Pullin analysis of conditional probabilities in quantum mechanics.
\bigskip

\noindent {\bf Acknowledgements:} R. B. and K. F. wants to thank Abhay Ashtekar, Jerzy  Lewandowski and Carlo Rovelli for useful hints and interesting discussions.

\end{document}